\documentclass[
    ,final            
    ,numberedheadings 
    ,cmfonts          
  ]
  {aipproc}

\layoutstyle{6x9}

\begin{document}

\title{Two Photon  Distribution Amplitudes}

\classification{12.38.Bx,14.70.Bh,12.20.Ds}
\keywords      {QCD factorization, Generalized distribution amplitude}

\author{M. El Beiyad\ }{
  address={Centre  de Physique Th{\'e}orique, \'Ecole Polytechnique, CNRS,
   91128 Palaiseau, France},
   altaddress={LPT, Universit\'e d'Orsay, CNRS, 91404 Orsay, France}
}

\author{B. Pire}{
  address={Centre  de Physique Th{\'e}orique, \'Ecole Polytechnique, CNRS,
   91128 Palaiseau, France}}

\author{L. Szymanowski}{
  address={Soltan Institute for Nuclear Studies, Warsaw, Poland}}

\author{S. Wallon}{
  address={ LPT, Universit\'e d'Orsay, CNRS, 91404 Orsay, France}}

\begin{abstract}
The factorization of the  amplitude of the process $\gamma^* \gamma \to \gamma \gamma$ in the low energy and high photon virtuality region
 is demonstrated at the Born order and in the leading logarithmic approximation.
The leading order two photon (generalized) distribution amplitudes  exhibit a characteristic  $\ln Q^2$
 behaviour and obey new inhomogeneous evolution equations.

\end{abstract}

\maketitle

\section{Introduction}
The pointlike coupling to quarks of the photon 
enables to calculate perturbatively part of its wave function. A twist expansion generates non-leading 
components of the photon distribution amplitude~\cite{Braun}, from which the lowest 
order one is chiral-odd and proportionnal to the magnetic susceptibility of the vacuum. 
The study of the two photon state is kinematically richer and is thus a most welcome 
theoretical laboratory for the study of exclusive hard reactions. 

The parton content of the photon has been the subject of many studies since the seminal paper
 by Witten~\cite{Witten}. A recent paper~\cite{FPS} extended the notion of 
 anomalous parton distribution in a photon   to the case of generalized parton distributions 
 (GPDs) used for the factorized description of the non-diagonal kinematics of deeply virtual
  Compton scattering (DVCS) on a real photon target, $\gamma^*(q) \gamma \to \gamma \gamma$, namely at
   large energy and small hadronic momentum transfer but large photon virtuality 
   ($Q^2=-q^2$). 

As for the  two meson case \cite{GDA} the two photon generalized distribution amplitudes
 describe the coupling of a quark antiquark (or gluon-gluon) 
pair to a pair of  photons, and are related by crossing to the photon GPDs.

We  study \cite{EPSW}  the scattering amplitude of the $\gamma^*(q) \gamma \to \gamma \gamma$
 process in the near threshold kinematics, namely at small $s$  and large $-t \sim Q^2$, at
  large $Q^2$ and in the leading order of the electromagnetic and strong couplings. This 
  enables us to define and calculate perturbatively the Born approximation of the  diphoton 
  GDAs.

This is reminiscent of, but quite different from, the
perturbative calculation of the rho rho GDA in terms of the rho
DA \cite{TDA-GDA}, for which both incoming photon were chosen to be 
hard in order to justify this factorization of the GDA.

\section{   $\gamma^*(q) \gamma(p_1) \to \gamma(q') \gamma(p_2)$ near threshold}

 Two photon production in  Compton scattering  on  a photon target 
\begin{equation}
\gamma^*(q) \gamma(q') \to \gamma(p_{1}) \gamma(p_2)
\label{dvcs}
\end{equation}
involves, at leading order in $\alpha_{em}$, and zeroth order in  $\alpha_{S}$  six "box"
Feynman diagrams with quarks in the loop.

Restricting  to the threshold kinematics where $W^2 = 
(p_{1}+p_{2})^2 = 0$ simplifies greatly the tensorial structure of the amplitude  while still preserving the richness
of the  skewedness ($\zeta$) dependence of GDAs. Our conventions for the kinematics are the following:

\begin{equation}\nonumber
q= p - \frac{Q^2}{s}n\ , ~~~~~~~~~~ q'= \frac{Q^2}{s}n\ ,
\end{equation}

\begin{equation}\nonumber
p_1=\zeta p\ , ~~~~~~~~~~ p_2= \bar \zeta p\ , ~~~~~~~~~~   \bar \zeta = 1-\zeta ,
\end{equation}
where $p$ and $n$ are two light-cone vectors and $2 p\cdot n =s $.
The momentum $l$ in the quark loop is parametrized as
\begin{equation}
l^\mu = z p^\mu +\beta n^\mu + l_T\ ,
\label{Sudakov}
\end{equation}
with $l_T^2=-{\mathbf l^2}$.
The  scattering amplitude is  written as
\begin{equation}
A = \epsilon_\mu\epsilon'_\nu{\epsilon_1}^*_\alpha{\epsilon^*_2}_\beta T^{\mu\nu\alpha\beta},
\end{equation}
where   the four photon polarization vectors are 
transverse with respect to Sudakov vectors $p$ and $n.$

The tensorial decomposition of $T^{\mu\nu\alpha\beta} (W=0)$ reads 
\begin{equation}
T^{\mu\nu\alpha\beta}  = \frac{1}{4}g^{\mu\nu}_Tg^{\alpha\beta}_T W_1+
\frac{1}{8}\left(g^{\mu\alpha}_Tg^{\nu\beta}_T 
+g^{\nu\alpha}_Tg^{\mu\beta}_T -g^{\mu\nu}_Tg^{\alpha\beta}_T \right)W_2
+ \frac{1}{4}\left(g^{\mu\alpha}_Tg^{\nu\beta}_T - g^{\mu\beta}_Tg^{\alpha\nu}_T\right)W_3\, ,
\end{equation}
and it involves three scalar functions $W_i$, $i=1,2,3$.

The integration over $l$ is performed as usual within the Sudakov representation, using
$$d^4l = \frac{s}{2} \, dz \, d\beta \, d^2 l_{T} \rightarrow \frac{\pi s}{2} \, dz \, d\beta \, d {\mathbf l^2}\,.$$
Let us  note that in order to interpret our result in terms of  factorized quantities, we will keep our expressions unintegrated with respect to $z$,
 the mass of the quark  playing the role of an infrared regulator.

One first integrates in $\beta$ using the Cauchy theorem. The propagators induce poles in the complex 
$\beta$-plane and the pole positions depend on the values of $z$ and $\zeta$.
The four poles lie all below the real axis for $z>1$ and lie all above the real 
axis for $z<0$, the only region where the amplitude may not vanish is   $1 > z > 0$. This leads to a natural interpretation
of $z$ as a partonic fraction of momentum.
One then identifies different regions defined from the relative values of $z$, $\zeta$ and $1-\zeta$.
This is reminiscent of the different  regions encountered in the kinematics of the generalized parton distributions $H(x,\xi,t)$, 
with the boundaries controlled by the relative values of $x$ and $\xi$, where $x$ and $\xi$ are the quantities related to our $z$ and $\zeta$ 
variables. For each of the six contributing diagrams the remaining integral over ${\mathbf l^2}$ contains a UV divergent 
part which cancels in their sum. 
  We get
\begin{eqnarray}
 W_{1}  &=& \frac{e_q^4N_C}{2\pi^2}\int_0^1dz\ (2z-1)\left[\frac{2z-\zeta}{z\bar{\zeta}}\theta(z-\zeta)+\frac{2z-1-\zeta}{\bar{z}\zeta}\theta(\zeta-z) \right. \nonumber \\
 &+&\left. \frac{2z-\bar{\zeta}}{z\zeta}\theta(z-\bar{\zeta})+\frac{2z-1-\bar{\zeta}}{\bar{z}\bar{\zeta}}
 \theta(\bar{\zeta}-z)\right]\log\frac{m^2}{Q^2}\; ,
 \label{W1}
\end{eqnarray}
\begin{equation}
 W_{2} =0 \; ,
\label{W2}
\end{equation}
and
\begin{eqnarray}
 W_{3}  &=& -\frac{e_q^4N_C}{2\pi^2}\int_0^1dz\ \left[\frac{\zeta}{z\bar{\zeta}}\theta(z-\zeta)-\frac{\bar{\zeta}}{\bar{z}\zeta}\theta(\zeta-z) \right. \nonumber \\
 &-&\left. \frac{\bar{\zeta}}{z\zeta}\theta(z-\bar{\zeta})+\frac{\zeta}{\bar{z}\bar{\zeta}}\theta(\bar{\zeta}-z)\right]\log\frac{m^2}{Q^2}.
 \label{W3}
\end{eqnarray}
Let us now interpret the results (\ref{W1}) and (\ref{W3}) from the 
point of view of QCD factorization based on the operator product expansion.

\section{ QCD factorization and the $\gamma \gamma$ GDA}

Let us consider two quark non local correlators on the light cone and their matrix elements 
between the vacuum and a diphoton state which define the diphoton GDA $\Phi_1$, 
\begin{equation}
\label{Fqa}
F^q = \int \frac{dy}{2\pi} e^{i(2z-1)\frac{y}{2}}\langle \gamma(p_{1})  \gamma(p_{2})| \bar q(\frac{-yN}{2})
 \gamma.N q(\frac{yN}{2})|0 \rangle = \frac{1}{2}g_\bot^{\mu\nu}\epsilon_\mu^*(p_1)\,\epsilon_\nu^*(p_2)\,\Phi_1(z,\zeta,0)
\end{equation}
where we denote $N= n/n.p$ and where we did not write explicitely 
the electromagnetic and the gluonic Wilson lines. We need also to define the matrix element
 of the  photonic correlator
\begin{equation}
\label{Fpa}
F^\gamma = \int \frac{dy}{2\pi} e^{i(2z-1)\frac{y}{2}}\langle \gamma(p_{1})  \gamma(p_{2})|F^{N\mu}(-\frac{y}{2}N)F_\mu^N(\frac{y}{2}N) |0 \rangle 
\end{equation}
where $F^{N \mu} = N_\nu F^{\nu \mu}$ , which mixes  with the quark correlator
 (\ref{Fqa}) although they are not of the same order in $\alpha_{em}$ \cite{Witten}.

We regulate through the usual dimensional regularization procedure
the UV divergent quark correlator matrix elements and obtain (with  $\frac{1}{\hat\epsilon} = \frac{1}{\epsilon} +\gamma_{E}-\log 4\pi$)
\begin{equation}
F^q= -\frac{N_C\,e_{q}^2}{4\pi^2} g_T^{\mu\nu}\epsilon^*_\mu (p_{1})\epsilon^*_\nu(p_{2}) 
 \left[\frac{1}{\hat\epsilon} + \log{m^2}\right] F(z,\zeta)\,,
\label{Fgam}
\end{equation}
with $ F(z,\zeta) =$
\begin{equation}
\label{Fz}
 \frac{\bar{z}(2z-\zeta)}{\bar{\zeta}}\theta(z-\zeta)+
 \frac{\bar{z}(2z-\bar{\zeta})}{\zeta}\theta(z-\bar{\zeta})+
 \frac{z(2z-1-\zeta)}{\zeta}\theta(\zeta-z)+\frac{z(2z-1-
 \bar{\zeta})}{\bar{\zeta}}\theta(\bar{\zeta}-z)\; . \nonumber
\end{equation}

\begin{figure}
\includegraphics[width=5cm]{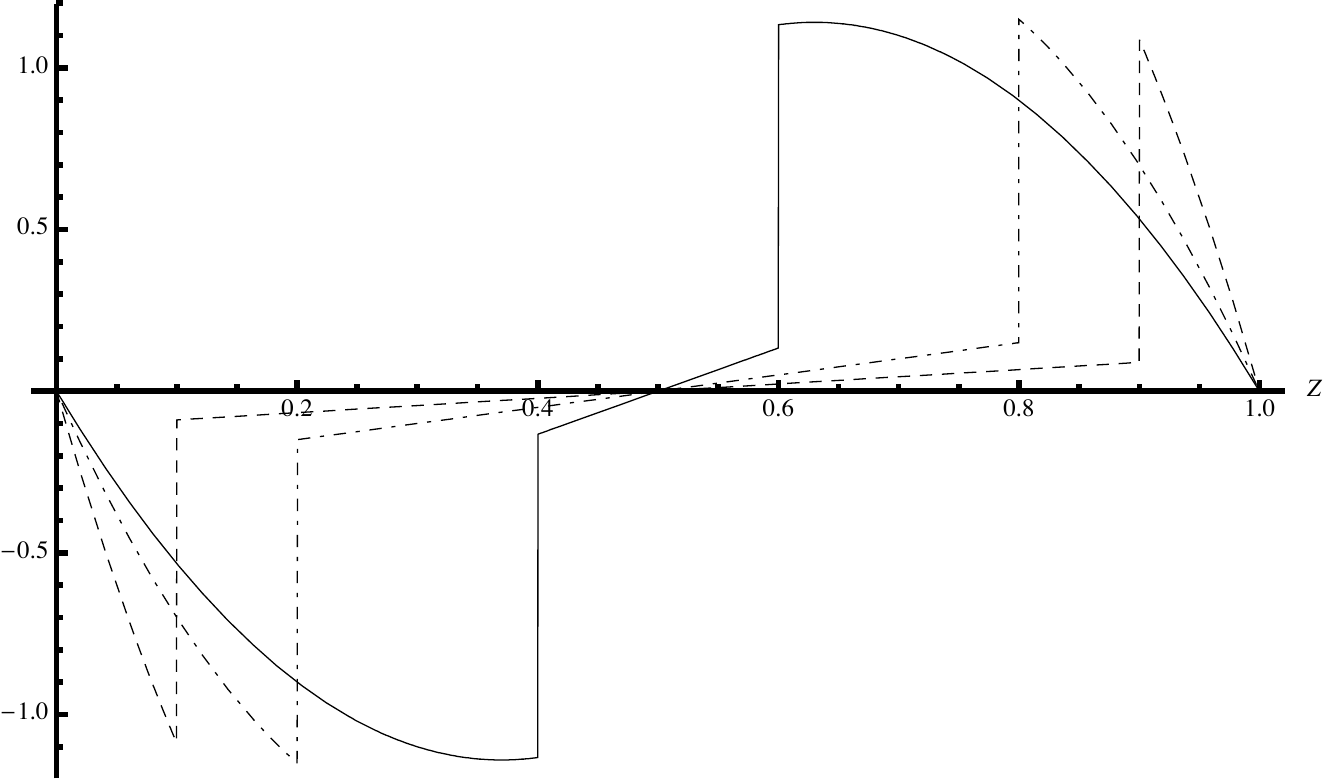}

\caption{\small The unpolarized anomalous diphoton GDA
 $\Phi_1^q/(N_C\,e_q^2/(2\pi^2)\,\log \frac{Q^2}{m^2})$ at Born order and 
 at threshold for $\zeta=0.1$ (dashed), $0.2$ (dash-dotted), $0.4$ (solid).}
\label{phi1}
\end{figure}

 The ultraviolet divergent parts
 are removed through the renormalization procedure (see for example \cite{Hill}) 
  involving quark and photon correlators    ($O^q$, $O^\gamma$) corresponding to  
$\bar q(-\frac{y}{2}N)\gamma.N q(\frac{y}{2}N)$ and $\ F^{N\mu}(-\frac{y}{2}N)F_\mu^N(\frac{y}{2}N) $.
 The renormalized operators are defined as :
   \begin{eqnarray}
\left(\begin{array}{c} O^q \\ O^\gamma \end{array}\right)_R = \left(\begin{array}{cc} Z_{qq} & Z_{q\gamma} \\ Z_{\gamma q} & Z_{\gamma\gamma} \end{array} \right)\left(\begin{array}{c} O^q \\ O^\gamma \end{array}\right).
\end{eqnarray}
   The matrix element of the renormalized quark-quark correlator is thus equal to
   \begin{eqnarray}
<\gamma(p_1)\gamma(p_2)|O_R^q|0> = Z_{qq}<\gamma(p_1)\gamma(p_2)|O^q|0>+Z_{q\gamma}<\gamma(p_1)\gamma(p_2)|O^\gamma|0>
\end{eqnarray}
with $Z_{qq} = 1+\mathcal{O}\left(\frac{e^2}{\hat{\epsilon}}\right)$. 
Since the matrix element $<\gamma(p_1)\gamma(p_2)|O^q|0>$ contains a UV divergence 
(Eqn. \ref{Fgam}) and since $<\gamma(p_1)\gamma(p_2)|O^\gamma|0>$ is UV finite and 
of order $\alpha_{em}^0$, one can absorb this divergence into the renormalization constant 
$Z_{q\gamma}$. The normalization of the renormalized correlator is fixed  with the help of 
the renormalization condition which is chosen as
  $<\gamma(p_1)\gamma(p_2)|O_R^q|0> = 0$  at the renormalization scale $M_R=m.$
    In this way the renormalized GDA  is equal to
    \begin{equation}
F^q_{R} 
 = -\frac{N_C\,e_{q}^2}{4\pi^2} g_T^{\mu\nu}\epsilon^*_\mu (p_{1})\epsilon^*_\nu(p_{2}) 
  \log{\frac{m^2}{M_{R}^2}} F(z,\zeta) \; .
\label{FRgam}
\end{equation}
from which we obtain - after identifying  the renormalized scale with  the factorization scale, $M_R=M_F$ -  that $\Phi^q_1(z,\zeta,0)=-\frac{N_ce_q^2}{2\pi^2}\log \frac{m^2}{M_F^2}F(z,\zeta)$. 
This expression together with the Born order coefficient function
 $C_V^q=e_q^2\left( \frac{1}{z}-\frac{1}{\bar z}  \right)$ leads to the quark 
 contribution to the $\gamma^* \gamma \to \gamma \gamma$ scattering amplitude
\begin{figure}
\includegraphics[width=5cm]{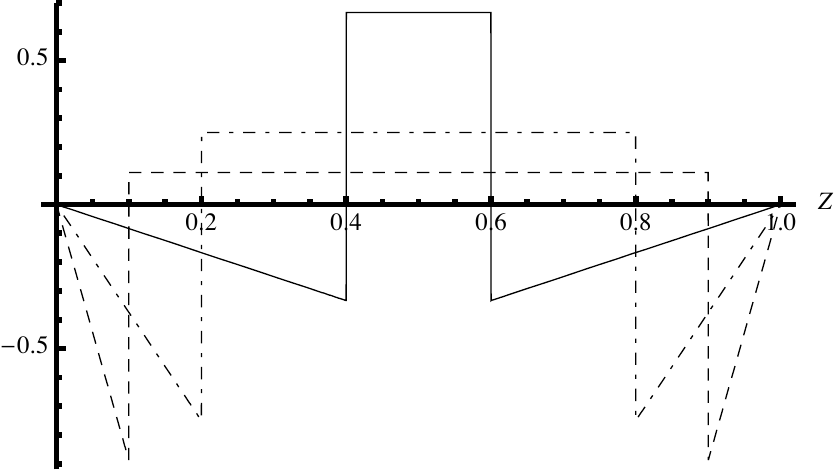}

\caption{\small The  polarized anomalous diphoton GDA 
$\Phi_3^q/(N_C\,e_q^2/(2\pi^2)\,\log \frac{Q^2}{m^2})$ at Born order and 
for $\zeta=0.1$ (dashed), $0.2$ (dash-dotted), $0.4$ (solid).}
\label{phi3}
\end{figure}
\begin{equation}
\label{W1q}
W_1^q = \int\limits_0^1\,dz\,C^q_V(z)\Phi^q_1(z,\zeta,0)\;.
\end{equation}
The contribution to $W_1$ in Eq.~(\ref{W1}) related to the photon operator involves a new
 coefficient function of order
$\alpha_{em}^2$ calculated at the factorization scale $M_F$ and convoluted with 
the zeroth order in $\alpha_{em}$ part of the photon GDA. This contribution coincides 
with the expression (\ref{W1}) in which the quark mass $m$ is replaced by the factorization
 scale $M_F$ playing now the role of infra-red cutoff. The sum of contributions related to the
  photon operator and to the quark operator (\ref{W1q}) reproduces then the result of direct
   calculations (\ref{W1}).
Let us note, that by choosing the factorization scale $M_F$ equal to the hard scale of our 
process $Q$, $M_F=Q$, the resulting $W_1$ comes only from the quark GDAcontribution.

We have thus demonstrated that it is legitimate to define the  Born order 
diphoton  GDAs at zero $W$ as
\begin{eqnarray}
\label{H1}
\Phi_1^q(z,\zeta,0) &=& \frac{N_C\,e_{q}^2}{2\pi^2} \log{\frac{Q^2}{m^2}}\left[\frac{\bar{z}(2z-\zeta)}{\bar{\zeta}}\theta(z-\zeta)+\frac{\bar{z}(2z-\bar{\zeta})}{\zeta}\theta(z-\bar{\zeta}) \right. \nonumber \\
&+&\left.\frac{z(2z-1-\zeta)}{\zeta}\theta(\zeta-z)+\frac{z(2z-1-\bar{\zeta})}{\bar{\zeta}}\theta(\bar{\zeta}-z)\right]
\end{eqnarray}

The same procedure can be applied to the axial vector correlator defining $\Phi_3$  
\begin{eqnarray}
\label{Fqaxial}
&&\int \frac{dy}{2\pi} e^{i(2z-1)\frac{y}{2}}\langle \gamma(p_{1})  \gamma(p_{2})| \bar q(-\frac{y}{2}N)
 \gamma.N \gamma_5 q(\frac{y}{2}N)|0 \rangle \\
 &&= -\frac{i}{2}\epsilon^{\mu \nu p N}\epsilon_\mu^*(p_1)\,\epsilon_\nu^*(p_2)\,\Phi_3(z,\zeta,0)\,.
\nonumber
\end{eqnarray}
We get
\begin{eqnarray}
\label{H3}
\Phi_3^q(z,\zeta,0) &=& \frac{N_C\,e_{q}^2}{2\pi^2} \log{\frac{Q^2}{m^2}}\left[\frac{\bar{z}\zeta}{\bar{\zeta}}\theta(z-\zeta)-\frac{\bar{z}\bar{\zeta}}{\zeta}\theta(z-\bar{\zeta}) \right. \nonumber \\
&-& \left. \frac{z\bar{\zeta}}{\zeta}\theta(\zeta-z)+\frac{z\zeta}{\bar{\zeta}}\theta(\bar{\zeta}-z)\right].
\end{eqnarray}
Since we focused on the logarithmic factors, we only obtained the {\em anomalous} part of 
these GDAs. Their $z-$ and $\zeta-$dependence  are shown on Figs. \ref{phi1} and 
\ref{phi3}.
Note that  they are discontinuous functions of $z$ at 
the points $z= \zeta$ and $z=\bar{\zeta}$.

\begin{theacknowledgments}
 We are grateful to Igor Anikin, Markus Diehl and Jean Philippe Lansberg for useful 
discussions and correspondance. 
This work is partly supported by the French-Polish scientific agreement Polonium, 
the Polish Grant 1 7294/R08/R09,  the ECO-NET program, contract 
18853PJ, the Joint Research Activity "Generalised Parton Distributions" of the european I3 program
Hadronic Physics, contract RII3-CT-2004-506078.   
\end{theacknowledgments}

\end{document}